\documentclass[journal]{IEEEtran}

\ifCLASSINFOpdf
\else
   \usepackage[dvips]{graphicx}
\fi
\usepackage{url}

\usepackage[utf8]{inputenc}
\usepackage[T1]{fontenc}
\usepackage{graphicx}
\usepackage{xcolor}
\usepackage{amssymb,amsmath}

\usepackage{tikz, pgfplots}
\pgfplotsset{compat=newest}
\usetikzlibrary{patterns}
\newlength\figurewidth

\newcommand{\qm}[1]{``#1''}  

\def\RR{\mathbb{R}}

\def\CC{\mathbb{C}}

\def\x{\vect{x}}
\def\u{\vect{u}}
\def\vv{\vect{v}}
\def\y{\vect{y}}
\def\z{\vect{z}}

\def\Mr{M_\mathrm{R}}

\def\Mh{M_\mathrm{H}}
\def\Ml{M_\mathrm{L}}

\def\One{\vect{1}}

\newcommand{\norm}[1]{\|#1\|}

\newcommand{\vect}[1]{\mathbf{#1}} 

\newcommand{\sdr}{\ensuremath{\mathrm{SDR}}}

\newcommand{\tc}{\ensuremath{\theta_\mathrm{c}}}

\definecolor{matlabCommentGreen}{RGB}{60,118,61}

\newcommand{\edt}[1]{{#1}}

\begin{document}

\title{Audio declipping performance enhancement\\via crossfading
}
\author{Pavel Záviška, Pavel Rajmic, and Ondřej Mokrý
\thanks{
The work of Pavel Záviška, Pavel Rajmic, and Ondřej Mokrý was supported by the Czech Science Foundation (GAČR) Project No.\,20-29009S.
}
\thanks{All three authors are with the Signal Processing Laboratory, 
Brno University of Technology, 61600 Brno, Czech Republic (e-mail: xzavis01@vutbr.cz; rajmic@feec.vutbr.cz; \edt{xmokry12@vutbr.cz}).}
}

\maketitle

\begin{abstract}
\edt{Some audio declipping methods produce waveforms that do not fully respect the physical process of clipping, which is why we refer to them as inconsistent.}
This letter reports what effect on perception it has if the solution by inconsistent methods is forced consistent by postprocessing.
We first propose a simple sample replacement method,
then we identify its main weaknesses and propose an improved variant.
The experiments show that the vast majority of inconsistent declipping methods significantly benefit from the proposed approach in terms of objective perceptual metrics.
In particular, we show that the SS\,PEW method based on social sparsity combined with the proposed method performs comparable to top methods from the consistent class,
but at a computational cost of one order of magnitude lower.
\end{abstract}

\begin{IEEEkeywords}
clipping, crossfade, declipping, optimization, sparsity, reliable samples
\end{IEEEkeywords}

\IEEEpeerreviewmaketitle

\section{Introduction}
Clipping is a~nonlinear distortion of signals, occurring in the case of lack of available dynamic range.
The peak values of a~signal are clipped (saturated).
In audio, this leads to undesirable, unpleasant perceptual artifacts.
Typically,
the best option is to avoid clipping beforehand.
\edt{In cases where clipping can no longer be prevented,}
there is a~need for a~means of estimating the original signal.
Such a~process is then called declipping.

There \edt{are numerous} 
methods aiming at declipping signals.
Since it is clearly an ill-conditioned task, any declipping method must build on some assumption about the behavior of a~signal.
In audio, which is the focus of this letter,
different methods are based on Bayesian modeling
\cite{FongGodsill2001:MonteCarlo},
on the autoregressive hypothesis
\cite{Dahimene2008_declipping},
on low-rank assumptions of matrices
\cite{BilenOzerovPerez2015:declipping.via.NMF,Takahashi2015:Block.Adaptive.Decipping.Null.Space},
but most of them are based on the sparsity of signal coefficients with respect to a~suitable time-frequency transform
\cite{Kitic2015:Sparsity.cosparsity.declipping,ZaviskaRajmicSchimmel2019:Psychoacoustics.l1.declipping}.
For more references, see the declipping survey \cite{ZaviskaRajmicOzerovRencker2021:Declipping.Survey} and a~recent overview \cite{GaultierKiticGribonvalBertin:Declipping2021}.

\begin{figure}
    \input{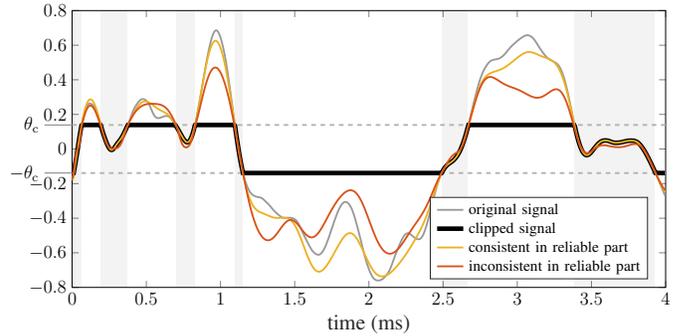}
    \caption{A~short piece of a~clipped signal, accompanied by its original 
    	version (\edt{unavailable in practice}) 
    	and by two types of reconstruction.
		The solution consistent in the reliable part is identical to the observed signal in that respective part,
		while the other is not.
				The scalars $-\tc,\tc$ are clipping thresholds, playing the role of the limits of the dynamic range.
        }
    \label{fig:waveforms.consistency}
\end{figure}

For
this letter, it will be
sufficient to distinguish between methods that produce declipping solutions that are consistent in the reliable part of the signal and methods that do not do so.
Here, the reliable part refers to samples that have not been affected by clipping, i.e., a set of samples that fitted 
within the prescribed dynamic range.
For illustration, Fig.\,\ref{fig:waveforms.consistency} shows an observed clipped signal,
together with the original and two declipping solutions;
one of them is consistent and the other is not.

The described declipping inconsistency is undeniably in conflict with the observation and with the clipping model.
On the other hand, inconsistent methods are typically faster, precisely because of the non-strict feasibility requirement.
What would happen if, for such methods, the observed reliable samples \edt{were} taken and put in their place in the current inconsistent solution?
This step would make the solution consistent and 
\edt{compatible} 
with the clipping model.
\edt{Certain} signal fidelity measures 
would improve automatically, 
\edt{such as the signal to distortion ratio (SDR),}
\begin{equation}
    \sdr (\u,\vv) = 20\log_{10} \frac{\norm{\u}_2}{\norm{\u-\vv}_2},
    \label{eq:SDR}
\end{equation}
\edt{when evaluated on the whole original and estimated signals as
$\sdr (\x_\mathrm{orig},\hat{\x})$.}

In audio, however, more attention is paid to the perceptual quality, and it is known that 
the SDR values seldom predict it well.
Thus, the interest of this letter lies in exploring how the \emph{perceptual} quality of the signal reconstruction changes when inconsistent samples in the solution are reconsidered with the help of reliable samples.

Section~\ref{sec:reliable.replacement} starts with the straightforward replacement,
and it shows that  such a~simple act indeed improves the objective perceptual metrics.
On the other hand, it also shows its weaknesses, leading to a~more \edt{inventive}, though still conceptually simple and computationally cheap
treatment.
Section~\ref{Sec:PEW} then discusses in greater detail a~particular inconsistent algorithm, SS\,PEW.
The section answers the question
whether the proposed techniques can improve this already successful algorithm such that it outperforms 
\edt{the state-of-the-art} 
audio declipping algorithms.

\section{What shall we do with the reliable samples?}
\label{sec:reliable.replacement}
In this section, we will examine
to what extent the signal reconstruction quality is affected when (re)using the reliable samples.
Note again that such a~postprocessing step makes sense only for declipping methods that produce solutions
inconsistent in the reliable part.
Specifically, we will follow up on the declipping survey
\cite{ZaviskaRajmicOzerovRencker2021:Declipping.Survey}
and work with the inconsistent methods included there:
The Constrained Orthogonal Matching Pursuit (C-OMP~\cite{Adler2011:Declipping}),
Plain and Perceptually-motivated Compressed Sensing L1 (CSL1, PCSL1~\cite{Defraene2013:Declipping.perceptual.compressed.sensing}), 
Parabola-weighted Compressed Sensing L1 (PWCSL1~\cite{ZaviskaRajmicOzerovRencker2021:Declipping.Survey}), 
Declipping with Empirical Wiener shrinkage
and Social Sparsity Declipping with Persistent Empirical Wiener (EW, PEW~\cite{SiedenburgKowalskiDoerfler2014:Audio.declip.social.sparsity}), \edt{and}
Dictionary Learning (DL~\cite{RenckerBachWangPlumbley2018:Consistent.dictionary.learning.LVA}).

Also, the audio database used in the experiments is identical to the one used in the declipping survey
\cite{ZaviskaRajmicOzerovRencker2021:Declipping.Survey}.
The corpus consists of 10~musical excerpts in mono, sampled at 44.1\,kHz
with a bit depth of 16\,bps and
with an approximate duration of 7~seconds.
The samples originally come from the EBU SQAM database\footnote{https://tech.ebu.ch/publications/sqamcd}\!
and they cover a~range of audio signal characteristics.
\edt{Per each signal, 7 degraded versions were obtained by (artificial) clipping}
using 7 variants of the symmetric thresholds $-\tc,\tc$.
\edt{These were computed}
\edt{according to}
prescribed input signal-to-distortion ratios,
\edt{defined as} $\sdr (\x_\mathrm{orig},\x_\mathrm{clip})$,
ranging from 1\,dB to 20\,dB.

For the perceptual quality assessment, the PEAQ (Perceptual Evaluation of Audio Quality)
metric was used.
The output of PEAQ is the Objective Difference Grade (ODG), which predicts the human rating of the difference between 
the~degraded
(or reconstructed)
\edt{and reference}
signals.
Possible values from $-4$ to $0$ correspond to the scale \qm{very annoying}---\qm{annoying}---\qm{slightly annoying}---\qm{perceptible, but not annoying}---\qm{imperceptible}.
We use the MATLAB code%
\footnote{\url{http://www-mmsp.ece.mcgill.ca/Documents/Software/}}
implemented according to the revised version of PEAQ (BS.1387-1)
\cite{Kabal2002:PEAQ}.

As another evaluation metric taking the human auditory system into account,
the PEMO-Q \cite{Huber:2006a} for Matlab%
\footnote{\url{https://www.hoertech.de/de/f-e-produkte/pemo-q.html}}
was used.
PEMO-Q computes the perceptual similarity measure (PSM), which can be mapped to the ODG score as used in the PEAQ.

\begin{figure}%
%
%
\definecolor{mycolor1}{rgb}{0.00000,0.44700,0.74100}%
\definecolor{mycolor2}{rgb}{0.8500, 0.3250, 0.0980}%
\definecolor{mycolor3}{rgb}{0.92900,0.69400,0.12500}%
%
\begin{tikzpicture}[scale=0.58]

\begin{axis}[%
width=5.4in, 
height=2.5in,
scale only axis,
xmin=0,
xmax=32,
xlabel style={font=\Large\color{white!15!black}},
xlabel={samples},
xticklabel style={font=\large},
minor x tick num=1,
xmajorgrids,
xminorgrids, 
minor x grid style={dotted, gray},
major x grid style={dotted, gray},
ymin=0,
ymax=0.14,
extra tick style={
        tick align=outside,
        tick pos=left,
        grid style={dotted,black}
    },
    extra y tick style={
        major tick length=2em
    },
extra y ticks = {0.0656043699524844},
extra y tick labels = {$\vspace{.2em}\tc$},
ylabel style={font=\Large\color{white!15!black}},
yticklabel style={font=\large},
yticklabels = {$$, $0$, $0.1$, $0.2$, $0.3$, $0.4$, $0.5$, $0.6$, $0.7$},
scaled ticks=false,
ylabel shift = -1em,
axis background/.style={fill=white},
legend style={at={(0.03,0.97)}, anchor=north west, legend cell align=left, align=left, draw=white!15!black},
every axis plot/.append style={very thick}
]
%
\addplot [color=white!70!black, dashed, forget plot]
  table[row sep=crcr]{%
0 0.0656043699524844\\
32 0.0656043699524844\\
};

%
%

\addplot [forget plot, draw=none, fill=black, fill opacity = 0.05] coordinates {
	(2.5, 1)
	(2.5, -1)
	(10.5, -1)
	(10.5, 1)
};
\addplot [forget plot, draw=none, fill=black, fill opacity = 0.05] coordinates {
	(18.5, 1)
	(18.5, -1)
	(24.5, -1)
	(24.5, 1)
};
\addplot [forget plot, draw=none, fill=black, fill opacity = 0.05] coordinates {
	(29.5, 1)
	(29.5, -1)
	(32.5, -1)
	(32.5, 1)
};


\addplot [color=white!60!black]
  table[row sep=crcr]{%
0	0.08795166015625\\
1	0.08734130859375\\
2	0.079345703125\\
3	0.0634765625\\
4	0.045074462890625\\
5	0.027191162109375\\
6	0.01312255859375\\
7	0.00482177734375\\
8	0.01220703125\\
9	0.031524658203125\\
10	0.0538330078125\\
11	0.080322265625\\
12	0.10906982421875\\
13	0.129730224609375\\
14	0.136260986328125\\
15	0.131072998046875\\
16	0.115020751953125\\
17	0.091522216796875\\
18	0.06640625\\
19	0.04443359375\\
20	0.027923583984375\\
21	0.021240234375\\
22	0.024383544921875\\
23	0.035858154296875\\
24	0.055389404296875\\
25	0.076690673828125\\
26	0.092529296875\\
27	0.1002197265625\\
28	0.096954345703125\\
29	0.082733154296875\\
30	0.06024169921875\\
31	0.03460693359375\\
32	0.0118408203125\\
};
\addlegendentry{original}


\addplot [color=mycolor1, line width = 3pt]
  table[row sep=crcr]{%
0	0.0670377220001737\\
1	0.0662337203925051\\
2	0.062443678094692\\
3	0.0551098322856203\\
4	0.0451596539944027\\
5	0.0349551559550016\\
6	0.0275145043061647\\
7	0.0252877503792019\\
8	0.0290790960637694\\
9	0.0377066392915415\\
10	0.0485886871897281\\
11	0.0589114906753318\\
12	0.0667254468877289\\
13	0.0713845776240429\\
14	0.0731650157169671\\
15	0.0724644206269227\\
16	0.0692605252053266\\
17	0.063233046757188\\
18	0.0544201209062374\\
19	0.0439380553317546\\
20	0.034216480509355\\
21	0.028337807027853\\
22	0.0285946078861229\\
23	0.035161028662755\\
24	0.0459390630297842\\
25	0.0576415527681796\\
26	0.0670355176451204\\
27	0.0714763526200244\\
28	0.0690775508278403\\
29	0.0592575931563836\\
30	0.0434525218165188\\
31	0.0250277732836302\\
32	0.00804152448230826\\
};
\addlegendentry{rec}

\addplot [color=mycolor2]
  table[row sep=crcr]{%
0	0.0670377220001737\\
1	0.0662337203925051\\
2	0.062443678094692\\
3	0.0634765625\\
4	0.045074462890625\\
5	0.027191162109375\\
6	0.01312255859375\\
7	0.00482177734375\\
8	0.01220703125\\
9	0.031524658203125\\
10	0.0538330078125\\
11	0.0589114906753318\\
12	0.0667254468877289\\
13	0.0713845776240429\\
14	0.0731650157169671\\
15	0.0724644206269227\\
16	0.0692605252053266\\
17	0.063233046757188\\
18	0.0544201209062374\\
19	0.04443359375\\
20	0.027923583984375\\
21	0.021240234375\\
22	0.024383544921875\\
23	0.035858154296875\\
24	0.055389404296875\\
25	0.0576415527681796\\
26	0.0670355176451204\\
27	0.0714763526200244\\
28	0.0690775508278403\\
29	0.0592575931563836\\
30	0.06024169921875\\
31	0.03460693359375\\
32	0.0118408203125\\
};
\addlegendentry{RR}

\addplot [color=mycolor3]
  table[row sep=crcr]{%
0	0.0670377220001737\\
1	0.0662337203925051\\
2	0.062443678094692\\
3	0.056335111557337\\
4	0.0451170584425139\\
5	0.0283281726835217\\
6	0.01312255859375\\
7	0.00482177734375\\
8	0.0146778879356671\\
9	0.0346156487473332\\
10	0.0493567001635748\\
11	0.0589114906753318\\
12	0.0667254468877289\\
13	0.0713845776240429\\
14	0.0731650157169671\\
15	0.0724644206269227\\
16	0.0692605252053266\\
17	0.063233046757188\\
18	0.0544201209062374\\
19	0.0440106252529374\\
20	0.031070032246865\\
21	0.0222796498250282\\
22	0.0250002408149873\\
23	0.035509591479815\\
24	0.0473230334660861\\
25	0.0576415527681796\\
26	0.0670355176451204\\
27	0.0714763526200244\\
28	0.0690775508278403\\
29	0.0592575931563836\\
30	0.0459112399218035\\
31	0.0298173534386901\\
32	0.0112844263200353\\
};
\addlegendentry{CR}

\end{axis}
\end{tikzpicture}%
	\caption{Demonstration of the replacement strategies on a~short piece of audio signal.
		Non-smooth transitions are present when the straightforward substitution (RR) is applied,
		but they are smoothed out when crossfading (CR) is applied.
		}%
	\label{fig:waveform.substitution}%
\end{figure}
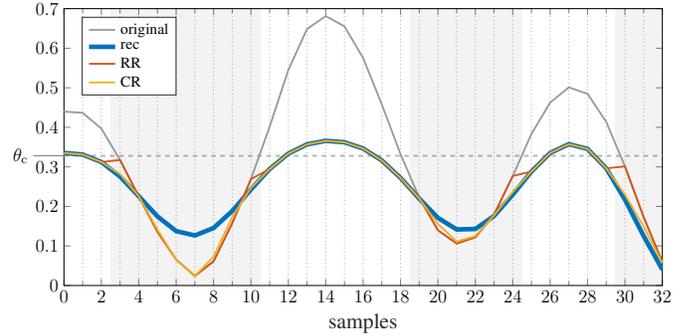

\subsection{Basic Replacement Strategy (RR as Replace Reliable)}
The basic approach has already been indicated:
the samples of the reconstructed signal are simply substituted with the reliable samples
in all parts where the signal was not clipped. 
Fig.\,\ref{fig:waveform.substitution} illustrates such an approach.
At the same time, this figure
reveals
the main, not unexpected problem of RR,
which is the risk of creating sharp transitions between the newly recreated reliable parts and the rest of the signal.
Perceptually, a~non-smooth phenomenon like this results in an undesirable occurrence of broadband spectral components.

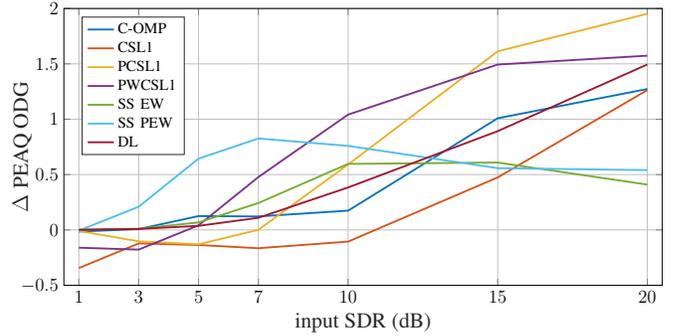
\begin{figure}%
%
%
\definecolor{mycolor1}{rgb}{0.00000,0.44700,0.74100}%
\definecolor{mycolor2}{rgb}{0.85000,0.32500,0.09800}%
\definecolor{mycolor3}{rgb}{0.92900,0.69400,0.12500}%
\definecolor{mycolor4}{rgb}{0.49400,0.18400,0.55600}%
\definecolor{mycolor5}{rgb}{0.46600,0.67400,0.18800}%
\definecolor{mycolor6}{rgb}{0.30100,0.74500,0.93300}%
\definecolor{mycolor7}{rgb}{0.63500,0.07800,0.18400}%
\begin{tikzpicture}[scale=0.58]

\begin{axis}[%
width=5.4in,
height=2.5in, 
at={(0.758in,0.481in)},
scale only axis,
xmin=0.5,
xmax=20.5,
xtick={ 1,  3,  5,  7, 10, 15, 20},
xlabel style={font=\color{white!15!black}},
xlabel={input SDR (dB)},
xticklabel style={font=\large},
xlabel style={font=\Large\color{white!15!black}},
ymin=-0.5,
ymax=2,
ylabel style={font=\color{white!15!black}},
ylabel={$\Delta$ PEAQ ODG},
ylabel style={font=\Large\color{white!15!black}},
yticklabel style={font=\large},
ylabel shift = -1.38em,
axis background/.style={fill=white},
xmajorgrids,
ymajorgrids,
legend style={at={(0.03,0.97)}, anchor=north west, legend cell align=left, align=left, draw=white!15!black},
every axis plot/.append style={very thick}
]

\addplot [color=mycolor1]
  table[row sep=crcr]{%
1	-0.0148206593008757\\
3	0.00763721694499595\\
5	0.125247252045489\\
7	0.12197774581007\\
10	0.174413961152851\\
15	1.00860490660886\\
20	1.27304961277126\\
};
\addlegendentry{C-OMP}

\addplot [color=mycolor2]
  table[row sep=crcr]{%
1	-0.344313943605183\\
3	-0.121634660173205\\
5	-0.136002424006678\\
7	-0.164999362452318\\
10	-0.105276700921001\\
15	0.474474095677232\\
20	1.26314054652945\\
};
\addlegendentry{CSL1}

\addplot [color=mycolor3]
  table[row sep=crcr]{%
1	-0.00955626950259001\\
3	-0.10267749216847\\
5	-0.129748549098016\\
7	0.0013608667069247\\
10	0.591398415145\\
15	1.61310089194514\\
20	1.9522968364042\\
};
\addlegendentry{PCSL1}

\addplot [color=mycolor4]
  table[row sep=crcr]{%
1	-0.160372166635567\\
3	-0.177487615913103\\
5	0.0416117387625634\\
7	0.477648601048797\\
10	1.04163598230139\\
15	1.49385971044228\\
20	1.57389866585091\\
};
\addlegendentry{PWCSL1}

\addplot [color=mycolor5]
  table[row sep=crcr]{%
1	0.00365439074042611\\
3	0.0111158577997746\\
5	0.0678248984475856\\
7	0.243202799993628\\
10	0.596439847086553\\
15	0.608679605987667\\
20	0.409746629803991\\
};
\addlegendentry{SS EW}

\addplot [color=mycolor6]
  table[row sep=crcr]{%
1	-0.00599310898506626\\
3	0.209935683438652\\
5	0.643847787192845\\
7	0.826041048811936\\
10	0.759150999629911\\
15	0.5579655129409\\
20	0.54035813804143\\
};
\addlegendentry{SS PEW}

\addplot [color=mycolor7]
  table[row sep=crcr]{%
1	0.00102179087969967\\
3	0.00904454972795117\\
5	0.0368830351761914\\
7	0.109805296817912\\
10	0.383776906977678\\
15	0.892334699511976\\
20	1.49447425005831\\
};
\addlegendentry{DL}

\end{axis}
\end{tikzpicture}%
\caption{Average ODG improvements using the basic (RR) strategy, compared to the initial declipping.
}%
\label{fig:rec_vs_RR}%
\end{figure}

Yet, the gain in the perceptual quality of the declipped audio obtained by the simple replacement strategy can outweigh the just described disadvantage,
as visible in Fig.~\ref{fig:rec_vs_RR}.
This figure shows the average PEAQ ODG improvement obtained using the basic replacement strategy.
The average is computed over the ODG values of individual excerpts.
The depicted values of $\Delta$\,PEAQ ODG suggest that the RR strategy is, for some declipping methods,
not suitable at low input SDR levels (i.e., a~low number of reliable samples).
\edt{Generally}, the improvement grows with increasing input SDR,
even up to two ODG grades in the case of the PCSL1 algorithm and 20\,dB input SDR.
Note that the replacement (and thus the improvement) comes essentially for free from the computational viewpoint.

\subsection{Advanced Strategy (CR as Crossfade Reliable)}
The
smarter strategy of enhancing the perceived quality 
\edt{stems from the simple one}, 
but now avoiding sharp jumps on the borders between the reliable and the clipped regions.
This is achieved by crossfading
the inconsistent declipping solution with the
observed signal
such that the reconstructed signal gradually blends into the reliable parts of the signal.

The crossfading transition can be performed either in the clipped part, in the reliable part, or in the middle affecting both parts.
Transition in the clipped part preserves all the reliable samples but \edt{is affected by the observed clipped  samples.} 
A~crossfade in the reliable part ensures the smoothest transition from the clipped to the reliable part, although some of the reliable samples are altered this way.

There are several types of 
crossfade used in practice.
In this paper, we examined the simple linear crossfade,
which is usually used for highly correlated material,
and the smooth 
crossfade,
whose curve follows
the squared sine function.

An important parameter of a~crossfade is the length of the crossfaded section, which determines the number of modified samples.
In the case of transition in the reliable part, the longer the transition is, the smoother one signal blends into the other;
however, more samples will then differ from the ground truth.
Hand in hand with specifying the crossfade length, 
\edt{it must be decided} how to treat segments that are shorter than the predefined length.
These can be either ignored (keeping the samples from the restored signal unaltered),
replaced using the RR strategy or the length of the crossfade can be shortened to fit the length of the processed segment.

Different setups for the CR strategy are thus available.
Experimenting with the possible combinations
showed convincingly that the transition \emph{in the reliable part}\/ produced the best perceptual results according to both PEAQ and \mbox{PEMO-Q}.
But in terms of the width and shape of the crossfades and of the way of treating the short segments,
the results vary according to the evaluation metric.
PEAQ seems to respond positively to the smooth crossfade and to ignoring the processing of shorter segments,
while PEMO-Q favors the linear transition and adaptive shortening of the short transitions.
Nonetheless, the differences between these setups are negligible
(up to 0.1 on the ODG scale).

\begin{figure*}
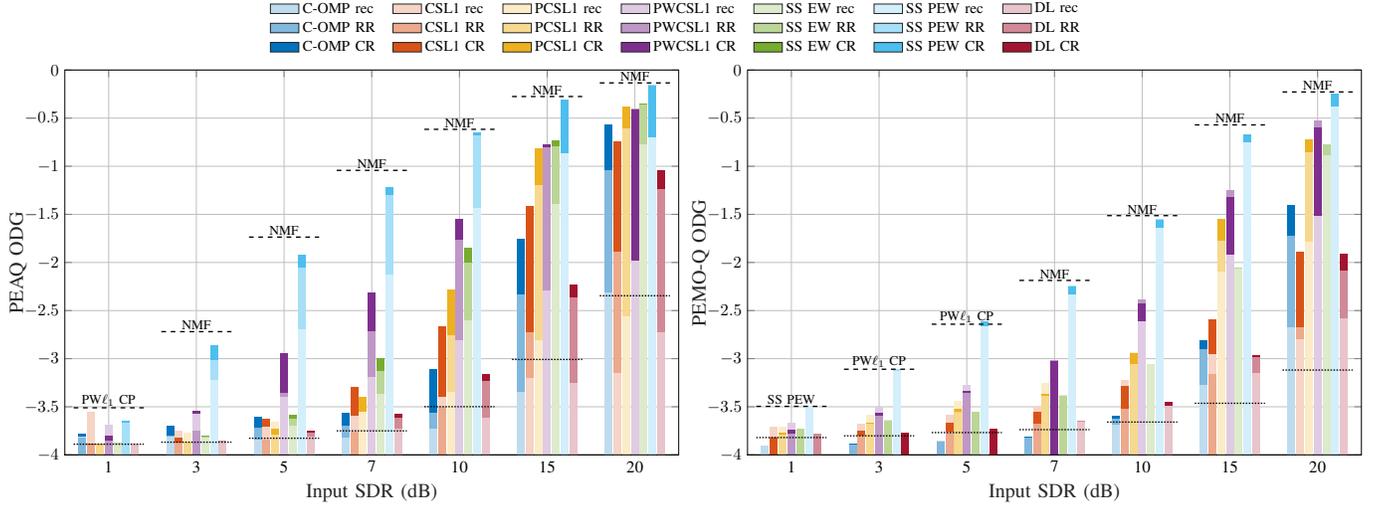

		\input{Results_PEAQ_sorted_colorReversed_spaceSaving.tex}
		\hspace{-0.75cm}
		\input{Results_PEMO-Q_sorted_colorReversed_spaceSaving.tex}
		\caption{Average PEAQ (left) and PEMO-Q (right) ODG values for inconsistent restoration (lightest color shade), RR strategy (medium shade) and CR strategy (darkest shade). 
		Each group of bars is crossed by a~horizontal dotted line; these mark the ODGs of the clipped signals.
		The dashed lines are present to indicate the best possible ODG results  achieved by a~method from the survey \cite{ZaviskaRajmicOzerovRencker2021:Declipping.Survey}.
		\vspace{-1em}
		}
	\label{fig:bar_results_all}
\end{figure*}

To allow a~concluding statement about the impact of the proposed CR strategy,
the same experiment as with the RR strategy was conducted.
As for the setup, the transition in the reliable part and an 8-sample-long smooth crossfade with adaptive shortening were used,
based on the above discussion.
Fig.~\ref{fig:bar_results_all} displays
the average PEAQ ODG and PEMO-Q ODG values. 
The individual declipping algorithms are
distinguished using different bar colors.
Within a~single bar, the lightest shade
represents the quality of the originally declipped, inconsistent signal.
The respective medium shade marks the results of the RR strategy, and finally, the darkest shade corresponds to CR.
In addition, the black dotted lines represent the average ODG value of the clipped signals, 
and the black dashed lines indicate the best ODG result obtained in the survey \cite{ZaviskaRajmicOzerovRencker2021:Declipping.Survey}.

The PEAQ results in Fig.\ \ref{fig:bar_results_all} (left) suggest a~significant improvement of the reconstruction quality
when the advanced crossfading method is applied,
especially at medium and high input SDRs.
\edt{The CR method always performs better or at least on par with the RR strategy.
However, in some cases of very harsh clipping (input SDR of 1 and 3 dB), both replacement methods can decrease the PEAQ score of the declipped signal.}
The results displayed in Fig.\ \ref{fig:bar_results_all} (right) indicate PEMO-Q being more conservative in comparison to PEAQ,
but even in this case the \edt{CR} strategy usually leads to an improvement, compared to \edt{RR}.

\begin{figure*} 
	\centering
	\definecolor{c1}{rgb}{0, 0.4470, 0.7410}
\definecolor{c2}{rgb}{0.8500, 0.3250, 0.0980}
\definecolor{c3}{rgb}{0.9290, 0.6940, 0.1250}
\definecolor{c4}{rgb}{0.4940, 0.1840, 0.5560}
\definecolor{c5}{rgb}{0.4660, 0.6740, 0.1880}
\definecolor{c6}{rgb}{0.3010, 0.7450, 0.9330}
\definecolor{c7}{rgb}{0.6350, 0.0780, 0.1840}

\begin{tikzpicture}[scale=0.63]  

\begin{axis}[%
hide axis,
xmin=0,
xmax=0,
ymin=0,
ymax=0,
legend style={at={(0,0)}, anchor=center, legend cell align=left, align=left, draw=white!15!black, draw=none, legend columns=3, transpose legend, font=\small},
]
\addlegendimage{ybar, fill=c1!25!white, draw=none, area legend}
\addlegendentry{C-OMP rec~~~}

\addlegendimage{ybar, fill=c1!50!white, draw=none, area legend}
\addlegendentry{C-OMP RR}

\addlegendimage{ybar, fill=c1, draw=none, area legend}
\addlegendentry{C-OMP CR}

\addlegendimage{ybar, fill=c2!25!white, draw=none, area legend}
\addlegendentry{CSL1 rec~~~}

\addlegendimage{ybar, fill=c2!50!white, draw=none, area legend}
\addlegendentry{CSL1 RR}

\addlegendimage{ybar, fill=c2, draw=none, area legend}
\addlegendentry{CSL1 CR}

\addlegendimage{ybar, fill=c3!25!white, draw=none, area legend}
\addlegendentry{PCSL1 rec~~~}

\addlegendimage{ybar, fill=c3!50!white, draw=none, area legend}
\addlegendentry{PCSL1 RR}

\addlegendimage{ybar, fill=c3, draw=none, area legend}
\addlegendentry{PCSL1 CR}

\addlegendimage{ybar, fill=c4!25!white, draw=none, area legend}
\addlegendentry{PWCSL1 rec~~~}

\addlegendimage{ybar, fill=c4!50!white, draw=none, area legend}
\addlegendentry{PWCSL1 RR}

\addlegendimage{ybar, fill=c4, draw=none, area legend}
\addlegendentry{PWCSL1 CR}

\addlegendimage{ybar, fill=c5!25!white, draw=none, area legend}
\addlegendentry{SS EW rec~~~}

\addlegendimage{ybar, fill=c5!50!white, draw=none, area legend}
\addlegendentry{SS EW RR }

\addlegendimage{ybar, fill=c5, draw=none, area legend}
\addlegendentry{SS EW CR}

\addlegendimage{ybar, fill=c6!25!white, draw=none, area legend}
\addlegendentry{SS PEW rec~~~}

\addlegendimage{ybar, fill=c6!50!white, draw=none, area legend}
\addlegendentry{SS PEW RR}

\addlegendimage{ybar, fill=c6, draw=none, area legend}
\addlegendentry{SS PEW CR}

\addlegendimage{ybar, fill=c7!25!white, draw=none, area legend}
\addlegendentry{DL rec}

\addlegendimage{ybar, fill=c7!50!white, draw=none, area legend}
\addlegendentry{DL RR}

\addlegendimage{ybar, fill=c7, draw=none, area legend}
\addlegendentry{DL CR}

\end{axis}
\end{tikzpicture}
	\vspace{-.5cm}
\end{figure*}		

\section{A closer look at SS\,PEW}
\label{Sec:PEW}

In the audio declipping survey \cite{ZaviskaRajmicOzerovRencker2021:Declipping.Survey},
the method based on the so-called social sparsity (SS) with persistent empirical Wiener (PEW) shrinkage
\cite{SiedenburgKowalskiDoerfler2014:Audio.declip.social.sparsity,siedenburg2012:wiener}
ranked high.
SS\,PEW was in fact the overall best-performing method in terms of the SDR.
In terms of PEMO-Q, 
SS\,PEW was also one of the
top-performing methods, 
being outperformed only slightly by two of the competitors,
namely the parabola-weighted $\ell_1$ minimization (PW$\ell_1$) and the non-negative matrix factorization (NMF).
However, PEAQ placed SS\,PEW a~bit deeper in the racing list.

PW$\ell_1$ and NMF
are actually fully consistent approaches,
and the NMF is computationally much more expensive in comparison with other declipping methods
\cite{ZaviskaRajmicOzerovRencker2021:Declipping.Survey}.
Since SS\,PEW produces generally inconsistent solutions, a~question arises as to
how much we can enhance the declipping quality of SS\,PEW by using the RR or CR methods.
How will it compare with PW$\ell_1$ and NMF then? 
Besides this, would it even be possible to reduce the computational time by computing fewer SS\,PEW iterations,
while still being able to reach the competitive quality of reconstruction with the help of the RR or CR strategies?

The algorithms based on social sparsity approximate the solution to the following optimization problem:
	\begin{multline}
    \label{eq:Social.sparsity.problem}
    \min_{\z}
    \left\{
        \frac{1}{2}\norm{\Mr D\z - \Mr\y }_2^2 
        + \frac{1}{2}\norm{h(\Mh D\z-\Mh\tc\One)}_2^2 + {} \right. \\
        \left. {} + \frac{1}{2}\norm{h(-\Ml D\z-\Ml\tc\One)}_2^2 + \lambda R(\z)
    \right\}.
  \end{multline}
The first term penalizes the $\ell_2$ error in the reliable region, which is captured by the mask operator $\Mr$.
\edt{The} vector $\z\in\CC^P$ represents coefficients approximating the signal $\y\in\RR^N$ with respect to a~selected time-frequency (TF) \edt{transform},
here a~synthesis operator
$D\colon\CC^P\to\RR^N$.
In the second and third terms, 
$\Mh$ and $\Ml$ are masks selecting only samples clipped from above and from below, respectively.
The deviation of the solution from the feasible (i.e., consistent) set in these clipped regions is penalized using the hinge function $h$, defined as
the identity for its negative arguments, and zero otherwise.
Finally, $R$ is a~sparsity-enforcing regularizer.
The authors of \cite{SiedenburgKowalskiDoerfler2014:Audio.declip.social.sparsity} suggest four types of shrinkage operator related to $R$ for use in practical algorithms.
The best-performing operator in audio declipping turned out to be the Persistent Empirical Wiener (PEW)
\cite{ZaviskaRajmicOzerovRencker2021:Declipping.Survey,GaultierKiticGribonvalBertin:Declipping2021,SiedenburgKowalskiDoerfler2014:Audio.declip.social.sparsity}:
\begin{equation}
		\mathcal{S}^{\textup{PEW}}_\lambda(z_{ft}) = z_{ft} \cdot
		\max \left(1-\frac{\lambda^2}{\norm{\mathcal{N}(z_{ft})}_{\edt{2}}^2},0 \right).
\label{eq:pew}
\end{equation}
The indexes $t$ and $f$ specify the location of a~coefficient in time and frequency, respectively,
and $\mathcal{N}(z_{ft})$ denotes the TF neighborhood of the coefficient
at location $(f,t)$.

The parameter $\lambda$ in
\eqref{eq:Social.sparsity.problem}
balances
sparsity and data fidelity.
Larger values of $\lambda$
lead
to a~solution of higher sparsity (meaning fewer coefficients) but also of greater deviance from the solution consistency, and vice versa.
To accelerate the overall convergence, the algorithm proposed in 
\cite{SiedenburgKowalskiDoerfler2014:Audio.declip.social.sparsity}
implements the adaptive restart strategy \cite{DonoghueCandes2013:Adaptive.restart}; 
the optimization actually starts with a~larger $\lambda$ and it is decreased after every few hundred iterations until the target value of $\lambda$ is reached.
This way, outer and inner iterations are recognized.

\begin{figure}%
%
%
\definecolor{mycolor1}{rgb}{0.00000,0.44700,0.74100}%
\definecolor{mycolor2}{rgb}{0.85000,0.32500,0.09800}%
\definecolor{mycolor3}{rgb}{0.92900,0.69400,0.12500}%
\definecolor{mycolor4}{rgb}{0.49400,0.18400,0.55600}%
\definecolor{mycolor5}{rgb}{0.46600,0.67400,0.18800}%
\definecolor{mycolor6}{rgb}{0.30100,0.74500,0.93300}%
\definecolor{mycolor7}{rgb}{0.63500,0.07800,0.18400}%
\begin{tikzpicture}[scale=0.58]

\begin{axis}[%
width=5.4in,
height=2.5in,
scale only axis,
xmin=0,
xmax=21,
xlabel style={font=\Large\color{white!15!black}},
xlabel={outer iterations},
xticklabel style={font=\large},
ymin=0,
ymax=40,
ylabel style={font=\Large\color{white!15!black}},
ylabel={SDR (dB)},
yticklabel style={font=\large},
axis background/.style={fill=white},
xmajorgrids,
ymajorgrids,
legend style={at={(0.03,0.97)}, anchor=north west, legend cell align=left, align=left, draw=white!15!black},
every axis plot/.append style={very thick}
]
\addlegendimage{empty legend}
\addlegendentry{\hspace{-.6cm}{Input SDR}}

\addplot [color=mycolor1]
  table[row sep=crcr]{%
1	0.265477564504369\\
2	0.447380510706182\\
3	0.746669624771014\\
4	1.14603765965557\\
5	1.65940916687077\\
6	2.31329275784189\\
7	3.12689338731045\\
8	4.10690078219334\\
9	5.35625350972151\\
10	6.81835327020041\\
11	8.44659282941962\\
12	10.0075183783754\\
13	11.2557163195307\\
14	12.1585016766481\\
15	12.6432347808357\\
16	12.9398416045556\\
17	13.0569531730327\\
18	13.1175492148625\\
19	13.1362841714013\\
20	13.1603037932826\\
};
\addlegendentry{1 dB}

\addplot [color=mycolor2]
  table[row sep=crcr]{%
1	2.76421486038395\\
2	3.79984880244089\\
3	4.97896644683136\\
4	6.24859679083814\\
5	7.60398974056684\\
6	9.12900276538425\\
7	10.7046286676276\\
8	12.2160105014162\\
9	13.5759124807942\\
10	14.6508011361318\\
11	15.6007214932916\\
12	16.2722364803204\\
13	16.7992755482632\\
14	17.2242483230415\\
15	17.4613577313012\\
16	17.6313048781656\\
17	17.7383981819527\\
18	17.838997461368\\
19	17.8845496324182\\
20	17.917018783926\\
};
\addlegendentry{3 dB}

\addplot [color=mycolor3]
  table[row sep=crcr]{%
1	5.56497013396283\\
2	7.01150030419985\\
3	8.53548480580598\\
4	10.105162728048\\
5	11.6893988544969\\
6	13.2391004463338\\
7	14.7937785983745\\
8	16.0821855404517\\
9	17.2485729255436\\
10	18.2300076915001\\
11	18.9451626880137\\
12	19.6208533310677\\
13	20.0886558132795\\
14	20.4622278315766\\
15	20.73577513664\\
16	20.8927084881893\\
17	21.0415123846045\\
18	21.1287614991624\\
19	21.1847851342066\\
20	21.212131020487\\
};
\addlegendentry{5 dB}

\addplot [color=mycolor4]
  table[row sep=crcr]{%
1	7.78285773603445\\
2	9.4651809313383\\
3	11.0690367218605\\
4	12.7656726072012\\
5	14.3853899152792\\
6	16.0068866634321\\
7	17.4934355559048\\
8	18.744600904118\\
9	19.8769167422993\\
10	20.7479117934375\\
11	21.4541049907885\\
12	22.0846903663864\\
13	22.6196645204518\\
14	22.9859243933064\\
15	23.2595766773883\\
16	23.4489982420107\\
17	23.5780259578751\\
18	23.6807461511022\\
19	23.7652062081957\\
20	23.8259841370972\\
};
\addlegendentry{7 dB}

\addplot [color=mycolor5]
  table[row sep=crcr]{%
1	10.2776082675922\\
2	12.0894707359526\\
3	13.8840268853341\\
4	15.6428310899843\\
5	17.3326847610834\\
6	18.9391370882251\\
7	20.3439188470294\\
8	21.55816768799\\
9	22.5544177250505\\
10	23.463547939366\\
11	24.2813377008277\\
12	24.9815626553332\\
13	25.5560107325519\\
14	26.0496497816059\\
15	26.395423890017\\
16	26.6432302830122\\
17	26.8489712203222\\
18	27.0235473495665\\
19	27.1297579797881\\
20	27.2202477890174\\
};
\addlegendentry{10 dB}

\addplot [color=mycolor6]
  table[row sep=crcr]{%
1	12.9009794635994\\
2	14.9020993935019\\
3	16.7984079909801\\
4	18.620098179579\\
5	20.480958565273\\
6	22.1155439953747\\
7	23.6101259571105\\
8	24.9963971404251\\
9	26.2492904900231\\
10	27.3823157332511\\
11	28.3321957065378\\
12	29.2534060396692\\
13	29.9928417484383\\
14	30.6402018696429\\
15	31.1420053934363\\
16	31.5212381689111\\
17	31.7697662890134\\
18	31.9665230843006\\
19	32.082918180493\\
20	32.152256582374\\
};
\addlegendentry{15 dB}

\addplot [color=mycolor7]
  table[row sep=crcr]{%
1	14.2084717858184\\
2	16.3107818229002\\
3	18.2921743880574\\
4	20.2583259312057\\
5	22.1596378995893\\
6	23.9544715120557\\
7	25.5536701999622\\
8	27.0169208238291\\
9	28.3943551066905\\
10	29.6777210692354\\
11	30.8410073422368\\
12	32.0232683145608\\
13	32.9442773279757\\
14	33.7183794452262\\
15	34.2312011738713\\
16	34.6576298841337\\
17	34.9770554377241\\
18	35.1789231673885\\
19	35.2849656655454\\
20	35.350338644211\\
};
\addlegendentry{20 dB}

\end{axis}

\end{tikzpicture}%
	\vspace{-1em}
	\caption{Average SDR in the clipped part in the course of iterations, with $\lambda = 10^{-4}$.
		\vspace{-.5em}
		}%
	\label{fig:sdr.course.clipped.4}%
\end{figure}
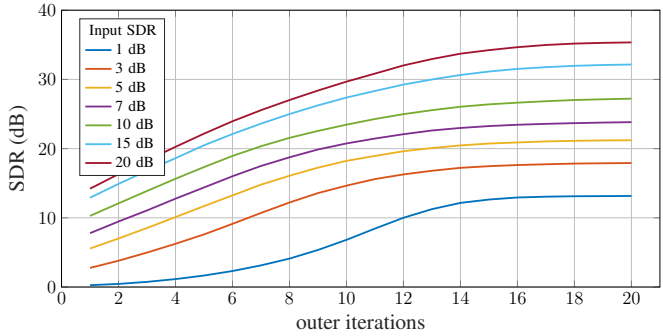

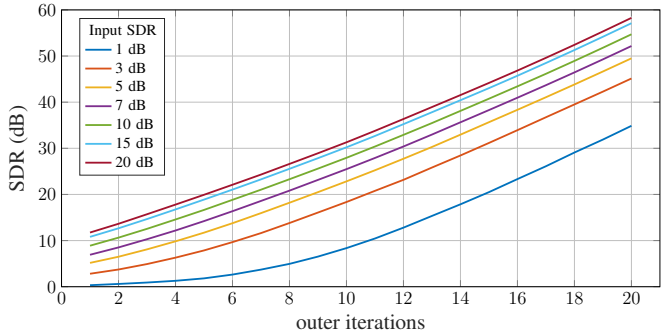
\begin{figure}%
%
%
\definecolor{mycolor1}{rgb}{0.00000,0.44700,0.74100}%
\definecolor{mycolor2}{rgb}{0.85000,0.32500,0.09800}%
\definecolor{mycolor3}{rgb}{0.92900,0.69400,0.12500}%
\definecolor{mycolor4}{rgb}{0.49400,0.18400,0.55600}%
\definecolor{mycolor5}{rgb}{0.46600,0.67400,0.18800}%
\definecolor{mycolor6}{rgb}{0.30100,0.74500,0.93300}%
\definecolor{mycolor7}{rgb}{0.63500,0.07800,0.18400}%
\begin{tikzpicture}[scale=0.58]

\begin{axis}[%
width=5.4in,
height=2.5in,
scale only axis,
xmin=0,
xmax=21,
xlabel style={font=\Large\color{white!15!black}},
xlabel={outer iterations},
xticklabel style={font=\large},
ymin=0,
ymax=60,
ylabel style={font=\Large\color{white!15!black}},
ylabel={SDR (dB)},
yticklabel style={font=\large},
axis background/.style={fill=white},
xmajorgrids,
ymajorgrids,
legend style={at={(0.03,0.97)}, anchor=north west, legend cell align=left, align=left, draw=white!15!black},
every axis plot/.append style={very thick}
]
\addlegendimage{empty legend}
\addlegendentry{\hspace{-.6cm}{Input SDR}}

\addplot [color=mycolor1]
  table[row sep=crcr]{%
1	0.32785780756201\\
2	0.598772145150951\\
3	0.905591354353246\\
4	1.28142915751359\\
5	1.80784812068307\\
6	2.6323290499635\\
7	3.70614735123926\\
8	4.95159989213162\\
9	6.52967398772882\\
10	8.36855272884\\
11	10.4507004041256\\
12	12.8045924719234\\
13	15.34220595216\\
14	17.8617745577719\\
15	20.4987787836722\\
16	23.3304682442318\\
17	26.113001485054\\
18	29.0859396564323\\
19	31.8956201293146\\
20	34.8732424054636\\
};
\addlegendentry{1 dB}

\addplot [color=mycolor2]
  table[row sep=crcr]{%
1	2.80972626374355\\
2	3.73076372394335\\
3	4.91003747571699\\
4	6.29983331211249\\
5	7.86270323432394\\
6	9.66759821887327\\
7	11.6215647289046\\
8	13.8090645506889\\
9	16.077357561482\\
10	18.3594330288891\\
11	20.7818615741632\\
12	23.1680786849158\\
13	25.8306568374793\\
14	28.4469950752956\\
15	31.1622064646205\\
16	33.9307645368297\\
17	36.7307028973703\\
18	39.507346622033\\
19	42.28861242414\\
20	45.1316666836769\\
};
\addlegendentry{3 dB}

\addplot [color=mycolor3]
  table[row sep=crcr]{%
1	5.16270902356884\\
2	6.49667397327287\\
3	8.09471468641368\\
4	9.83572457470487\\
5	11.7156400656901\\
6	13.7662587845024\\
7	15.9521680977613\\
8	18.2045662831028\\
9	20.4967561955077\\
10	22.8326388531051\\
11	25.2394284044075\\
12	27.7370940989142\\
13	30.289231081368\\
14	32.9587720156673\\
15	35.6593153509578\\
16	38.3435350427216\\
17	41.0688955986392\\
18	43.8351867417031\\
19	46.6495974845472\\
20	49.5025816255626\\
};
\addlegendentry{5 dB}

\addplot [color=mycolor4]
  table[row sep=crcr]{%
1	6.92426696537881\\
2	8.51609254776141\\
3	10.2958451389015\\
4	12.182717608755\\
5	14.2198511605365\\
6	16.3662823686317\\
7	18.5901084820996\\
8	20.8248350004827\\
9	23.1515338443008\\
10	25.4827070827665\\
11	27.9077718997561\\
12	30.3943263548146\\
13	32.9801078968679\\
14	35.6387753992349\\
15	38.2895606610352\\
16	40.9666722570398\\
17	43.6805991761686\\
18	46.4488870145933\\
19	49.2827006947779\\
20	52.1663204191131\\
};
\addlegendentry{7 dB}

\addplot [color=mycolor5]
  table[row sep=crcr]{%
1	8.88824982453955\\
2	10.6388385828269\\
3	12.5450144027623\\
4	14.5719563132886\\
5	16.6427552862491\\
6	18.8355609426099\\
7	21.0317965791674\\
8	23.2972822074124\\
9	25.5971439781426\\
10	27.9426096350787\\
11	30.3852179536508\\
12	32.915304745577\\
13	35.4716805536012\\
14	38.1088559222707\\
15	40.7508612378768\\
16	43.4303857132048\\
17	46.1491050336273\\
18	48.9427314646803\\
19	51.7953530505599\\
20	54.7084867144809\\
};
\addlegendentry{10 dB}

\addplot [color=mycolor6]
  table[row sep=crcr]{%
1	10.8064981865429\\
2	12.6625052055804\\
3	14.6642425532294\\
4	16.7510380461059\\
5	18.8833539646775\\
6	21.0522293925404\\
7	23.2664730212615\\
8	25.5548137845942\\
9	27.8612871199347\\
10	30.2152960288534\\
11	32.6742580113283\\
12	35.2344147387344\\
13	37.8209428603425\\
14	40.4227908015008\\
15	43.0419554324733\\
16	45.7294199812448\\
17	48.4902467577051\\
18	51.2964245092378\\
19	54.1817079524917\\
20	57.1088372352926\\
};
\addlegendentry{15 dB}

\addplot [color=mycolor7]
  table[row sep=crcr]{%
1	11.7424670245958\\
2	13.6434893169693\\
3	15.7077397318044\\
4	17.8012653924083\\
5	19.9246492260815\\
6	22.1014148121754\\
7	24.3069928846294\\
8	26.6164545338301\\
9	28.927911450676\\
10	31.2992434086412\\
11	33.7861741530275\\
12	36.350783652701\\
13	38.9421498559079\\
14	41.5376593296929\\
15	44.1693140122116\\
16	46.8493403564054\\
17	49.6113340692917\\
18	52.4288091676365\\
19	55.3287340180391\\
20	58.2685139387092\\
};
\addlegendentry{20 dB}

\end{axis}

\end{tikzpicture}%
	\vspace{-1em}
	\caption{Average SDR in the reliable part in the course of iterations, with $\lambda = 10^{-4}$.
		\vspace{-.5em}
		}%
	\label{fig:sdr.course.reliable.4}%
\end{figure}

Figs.\ \ref{fig:sdr.course.clipped.4} and \ref{fig:sdr.course.reliable.4} demonstrate 
the convergence of the SDR computed in the clipped and reliable parts, respectively.
The plots show that while the SDR continues to grow in the reliable part with the increasing number of iterations,
the SDR in the clipped \edt{part} stabilizes after reaching a~certain value.
The pictures show SDR for the particular case when $\lambda = 10^{-4}$, which is in agreement with the setup from 
\cite{ZaviskaRajmicOzerovRencker2021:Declipping.Survey},
but such behavior has also been verified for $\lambda = 10^{-5}$.
In other words, from a~certain point on,
iterations only minimize the difference in the reliable samples.
This observation supports the above-proposed idea of terminating the iterations of SS\,PEW earlier and applying the RR/CR postprocessing.

\begin{figure}%
%
%
\definecolor{mycolor1}{rgb}{0.00000,0.44700,0.74100}%
\definecolor{mycolor2}{rgb}{0.85000,0.32500,0.09800}%
\definecolor{mycolor3}{rgb}{0.92900,0.69400,0.12500}%
\definecolor{mycolor4}{rgb}{0.49400,0.18400,0.55600}%
\definecolor{mycolor5}{rgb}{0.46600,0.67400,0.18800}%
\definecolor{mycolor6}{rgb}{0.30100,0.74500,0.93300}%
\definecolor{mycolor7}{rgb}{0.63500,0.07800,0.18400}%
\begin{tikzpicture}[scale=0.58]

\begin{axis}[%
width=5.4in,
height=2.5in,
scale only axis,
xmin=0,
xmax=21,
xlabel style={font=\Large\color{white!15!black}},
xlabel={outer iterations},
xticklabel style={font=\large},
ymin=-4,
ymax=0,
ylabel style={font=\Large\color{white!15!black}},
ylabel={PEAQ ODG},
ylabel shift=-1pt,
yticklabel style={font=\large},
axis background/.style={fill=white},
xmajorgrids, 
ymajorgrids,
legend style={at={(0.03,0.97)}, anchor=north west, legend cell align=left, align=left, draw=white!15!black},
every axis plot/.append style={very thick}
]
\addlegendimage{empty legend}
\addlegendentry{\hspace{-.6cm}{Input SDR}}

\addplot [color=mycolor1]
  table[row sep=crcr]{%
1	-3.87733600099094\\
2	-3.87881364108794\\
3	-3.8866211740035\\
4	-3.89281165102816\\
5	-3.89342009647433\\
6	-3.88826043870348\\
7	-3.8819782960975\\
8	-3.88272043336054\\
9	-3.86671927609879\\
10	-3.85344314421843\\
11	-3.83462436659101\\
12	-3.79186097321519\\
13	-3.74057045441117\\
14	-3.69403471046461\\
15	-3.63312763794341\\
16	-3.58892711150366\\
17	-3.58292799728235\\
18	-3.59105701810704\\
19	-3.60782072182208\\
20	-3.64279978387795\\
};
\addlegendentry{1 dB}

\addplot [color=mycolor2]
  table[row sep=crcr]{%
1	-3.87029849766573\\
2	-3.87153134404781\\
3	-3.87069771403518\\
4	-3.86866996294457\\
5	-3.86600028991274\\
6	-3.85857185988087\\
7	-3.84935087673391\\
8	-3.82619026577166\\
9	-3.78338043189122\\
10	-3.71494942602389\\
11	-3.61505824839629\\
12	-3.48159050095323\\
13	-3.32175874430056\\
14	-3.18920724275115\\
15	-3.06434050265503\\
16	-2.98211672443265\\
17	-2.89899998198966\\
18	-2.85466856625144\\
19	-2.85637806875819\\
20	-2.86469714824781\\
};
\addlegendentry{3 dB}

\addplot [color=mycolor3]
  table[row sep=crcr]{%
1	-3.85289154295184\\
2	-3.84790183085109\\
3	-3.84022975880461\\
4	-3.82260021419747\\
5	-3.79464494196834\\
6	-3.75204398281094\\
7	-3.68059717265882\\
8	-3.57239861008312\\
9	-3.43072741037292\\
10	-3.21030337437824\\
11	-2.94749056394609\\
12	-2.67377382803351\\
13	-2.44742730978569\\
14	-2.27714608164597\\
15	-2.1510788019833\\
16	-2.06318072193995\\
17	-1.98903116935866\\
18	-1.95008046742817\\
19	-1.90391157551485\\
20	-1.91547332842722\\
};
\addlegendentry{5 dB}

\addplot [color=mycolor4]
  table[row sep=crcr]{%
1	-3.8101610463519\\
2	-3.77767846234344\\
3	-3.72588916909581\\
4	-3.65319721376084\\
5	-3.56750745667899\\
6	-3.46050335352518\\
7	-3.31062406034851\\
8	-3.0973066564955\\
9	-2.77418172451245\\
10	-2.4347011331608\\
11	-2.15531750375524\\
12	-1.89950433275169\\
13	-1.70197275283114\\
14	-1.55898978078939\\
15	-1.44919641638285\\
16	-1.35720161413364\\
17	-1.29639485596615\\
18	-1.25829152297805\\
19	-1.23043401265148\\
20	-1.21692213455293\\
};
\addlegendentry{7 dB}

\addplot [color=mycolor5]
  table[row sep=crcr]{%
1	-3.61524043798285\\
2	-3.50363224536525\\
3	-3.38025796897412\\
4	-3.23072607155757\\
5	-3.07014909017555\\
6	-2.84951603317808\\
7	-2.53840443230606\\
8	-2.1660603137673\\
9	-1.8419568094185\\
10	-1.5784083145153\\
11	-1.32782066392282\\
12	-1.13311983671778\\
13	-0.980914750080348\\
14	-0.876572110950288\\
15	-0.803956642748756\\
16	-0.748286486609937\\
17	-0.707782233369195\\
18	-0.684106644502868\\
19	-0.662617692585571\\
20	-0.652571390539993\\
};
\addlegendentry{10 dB}

\addplot [color=mycolor6]
  table[row sep=crcr]{%
1	-2.96561714425757\\
2	-2.75985959996494\\
3	-2.52327959557788\\
4	-2.26324782516876\\
5	-1.96234761436511\\
6	-1.65306227191059\\
7	-1.36947463677892\\
8	-1.13708929814927\\
9	-0.956224910021375\\
10	-0.797180905092729\\
11	-0.650257480645371\\
12	-0.542425246014226\\
13	-0.468139876184933\\
14	-0.415200467442008\\
15	-0.382268191119552\\
16	-0.356658461361985\\
17	-0.336676396580596\\
18	-0.321721925668254\\
19	-0.311966866613354\\
20	-0.304926380760084\\
};
\addlegendentry{15 dB}

\addplot [color=mycolor7]
  table[row sep=crcr]{%
1	-2.21411714974241\\
2	-1.96069355368267\\
3	-1.68504702090496\\
4	-1.3712446095981\\
5	-1.12417106324999\\
6	-0.914848912533489\\
7	-0.75214266109908\\
8	-0.622535542263038\\
9	-0.510147404156067\\
10	-0.424379445654694\\
11	-0.357085318301836\\
12	-0.304935986002142\\
13	-0.259603850087225\\
14	-0.224812944850003\\
15	-0.203854398166434\\
16	-0.193403661238674\\
17	-0.183627667702114\\
18	-0.177425599241425\\
19	-0.169840482270325\\
20	-0.164026299747569\\
};
\addlegendentry{20 dB}

\addplot [color=mycolor1, dotted, forget plot]
  table[row sep=crcr]{%
1	-3.66538988745506\\
2	-3.66538988745506\\
3	-3.66538988745506\\
4	-3.66538988745506\\
5	-3.66538988745506\\
6	-3.66538988745506\\
7	-3.66538988745506\\
8	-3.66538988745506\\
9	-3.66538988745506\\
10	-3.66538988745506\\
11	-3.66538988745506\\
12	-3.66538988745506\\
13	-3.66538988745506\\
14	-3.66538988745506\\
15	-3.66538988745506\\
16	-3.66538988745506\\
17	-3.66538988745506\\
18	-3.66538988745506\\
19	-3.66538988745506\\
20	-3.66538988745506\\
};
\addplot [color=mycolor2, dotted, forget plot]
  table[row sep=crcr]{%
1	-3.22527776230266\\
2	-3.22527776230266\\
3	-3.22527776230266\\
4	-3.22527776230266\\
5	-3.22527776230266\\
6	-3.22527776230266\\
7	-3.22527776230266\\
8	-3.22527776230266\\
9	-3.22527776230266\\
10	-3.22527776230266\\
11	-3.22527776230266\\
12	-3.22527776230266\\
13	-3.22527776230266\\
14	-3.22527776230266\\
15	-3.22527776230266\\
16	-3.22527776230266\\
17	-3.22527776230266\\
18	-3.22527776230266\\
19	-3.22527776230266\\
20	-3.22527776230266\\
};
\addplot [color=mycolor3, dotted, forget plot]
  table[row sep=crcr]{%
1	-2.69887524453718\\
2	-2.69887524453718\\
3	-2.69887524453718\\
4	-2.69887524453718\\
5	-2.69887524453718\\
6	-2.69887524453718\\
7	-2.69887524453718\\
8	-2.69887524453718\\
9	-2.69887524453718\\
10	-2.69887524453718\\
11	-2.69887524453718\\
12	-2.69887524453718\\
13	-2.69887524453718\\
14	-2.69887524453718\\
15	-2.69887524453718\\
16	-2.69887524453718\\
17	-2.69887524453718\\
18	-2.69887524453718\\
19	-2.69887524453718\\
20	-2.69887524453718\\
};
\addplot [color=mycolor4, dotted, forget plot]
  table[row sep=crcr]{%
1	-2.1253649664676\\
2	-2.1253649664676\\
3	-2.1253649664676\\
4	-2.1253649664676\\
5	-2.1253649664676\\
6	-2.1253649664676\\
7	-2.1253649664676\\
8	-2.1253649664676\\
9	-2.1253649664676\\
10	-2.1253649664676\\
11	-2.1253649664676\\
12	-2.1253649664676\\
13	-2.1253649664676\\
14	-2.1253649664676\\
15	-2.1253649664676\\
16	-2.1253649664676\\
17	-2.1253649664676\\
18	-2.1253649664676\\
19	-2.1253649664676\\
20	-2.1253649664676\\
};
\addplot [color=mycolor5, dotted, forget plot]
  table[row sep=crcr]{%
1	-1.43581379450443\\
2	-1.43581379450443\\
3	-1.43581379450443\\
4	-1.43581379450443\\
5	-1.43581379450443\\
6	-1.43581379450443\\
7	-1.43581379450443\\
8	-1.43581379450443\\
9	-1.43581379450443\\
10	-1.43581379450443\\
11	-1.43581379450443\\
12	-1.43581379450443\\
13	-1.43581379450443\\
14	-1.43581379450443\\
15	-1.43581379450443\\
16	-1.43581379450443\\
17	-1.43581379450443\\
18	-1.43581379450443\\
19	-1.43581379450443\\
20	-1.43581379450443\\
};
\addplot [color=mycolor6, dotted, forget plot]
  table[row sep=crcr]{%
1	-0.861468452213424\\
2	-0.861468452213424\\
3	-0.861468452213424\\
4	-0.861468452213424\\
5	-0.861468452213424\\
6	-0.861468452213424\\
7	-0.861468452213424\\
8	-0.861468452213424\\
9	-0.861468452213424\\
10	-0.861468452213424\\
11	-0.861468452213424\\
12	-0.861468452213424\\
13	-0.861468452213424\\
14	-0.861468452213424\\
15	-0.861468452213424\\
16	-0.861468452213424\\
17	-0.861468452213424\\
18	-0.861468452213424\\
19	-0.861468452213424\\
20	-0.861468452213424\\
};
\addplot [color=mycolor7, dotted, forget plot]
  table[row sep=crcr]{%
1	-0.697860454489953\\
2	-0.697860454489953\\
3	-0.697860454489953\\
4	-0.697860454489953\\
5	-0.697860454489953\\
6	-0.697860454489953\\
7	-0.697860454489953\\
8	-0.697860454489953\\
9	-0.697860454489953\\
10	-0.697860454489953\\
11	-0.697860454489953\\
12	-0.697860454489953\\
13	-0.697860454489953\\
14	-0.697860454489953\\
15	-0.697860454489953\\
16	-0.697860454489953\\
17	-0.697860454489953\\
18	-0.697860454489953\\
19	-0.697860454489953\\
20	-0.697860454489953\\
};
\end{axis}
\end{tikzpicture}%
	\vspace{-1em}
	\caption{Average course of PEAQ ODG values for SS\,PEW followed by CR through iterations.
	The dotted lines indicate the best ODG score achievable by the pure SS\,PEW (after 20 iterations with $\lambda=10^{-4}$).
	\vspace{-.5em}}%
	\label{fig:peaq.course}%
\end{figure}
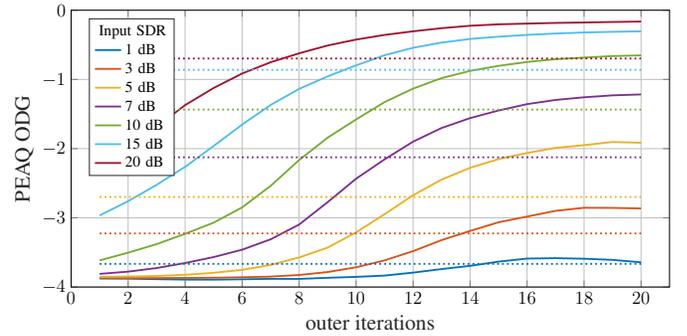

To examine if the presented hypotheses are true, 
we have performed an experiment:
The SS\,PEW was run for 20 (outer) iterations,
but after each iteration, the PEAQ and \mbox{PEMO-Q} ODG were computed for the current declipping solution, with the CR applied.
Fig.~\ref{fig:peaq.course} not only shows that the CR strategy raises the limit of the achievable ODG via SS\,PEW
(this fact was also visible in Fig.~\ref{fig:bar_results_all}).
More importantly, it shows that by applying the CR, results perceptually similar to the performance of a~pure SS\,PEW can be reached with significantly fewer iterations
(the savings range from one third to two thirds of computational power, depending on the input SDR).

\vspace{-.18em}
		
\section{Conclusion}
We have shown that results of inconsistent audio declipping methods can perceptually be significantly improved 
using a~computationally cheap technique, based on crossfading the regions of the signal.
In particular, the SS\,PEW method postprocessed with the proposed strategy becomes the top-performing inconsistent method and its quality almost 
reaches the quality of the best-performing consistent declipper
(NMF in the majority of cases).
This is especially attractive, considering that the NMF method is at least 15 times slower than the SS\,PEW.
In addition, we have shown that the SS\,PEW can be accelerated about twice without sacrificing perceptual quality.

Since this letter can be viewed as a~follow-up to the audio declipping survey \cite{ZaviskaRajmicOzerovRencker2021:Declipping.Survey},
the results are appended to the website
\begin{center}
\url{https://rajmic.github.io/declipping2020}
\end{center}
where the audio samples are directly playable.
Also the Matlab source codes are available through the link on that website.

\clearpage
\newcommand{\noopsort}[1]{} \newcommand{\printfirst}[2]{#1}
  \newcommand{\singleletter}[1]{#1} \newcommand{\switchargs}[2]{#2#1}


\begin{thebibliography}{10}
\providecommand{\url}[1]{#1}
\csname url@samestyle\endcsname
\providecommand{\newblock}{\relax}
\providecommand{\bibinfo}[2]{#2}
\providecommand{\BIBentrySTDinterwordspacing}{\spaceskip=0pt\relax}
\providecommand{\BIBentryALTinterwordstretchfactor}{4}
\providecommand{\BIBentryALTinterwordspacing}{\spaceskip=\fontdimen2\font plus
\BIBentryALTinterwordstretchfactor\fontdimen3\font minus
  \fontdimen4\font\relax}
\providecommand{\BIBforeignlanguage}[2]{{%
\expandafter\ifx\csname l@#1\endcsname\relax
\typeout{** WARNING: IEEEtran.bst: No hyphenation pattern has been}%
\typeout{** loaded for the language `#1'. Using the pattern for}%
\typeout{** the default language instead.}%
\else
\language=\csname l@#1\endcsname
\fi
#2}}
\providecommand{\BIBdecl}{\relax}
\BIBdecl

\bibitem{FongGodsill2001:MonteCarlo}
W.~Fong and S.~Godsill, ``Monte carlo smoothing for non-linearly distorted
  signals,'' in \emph{2001 IEEE International Conference on Acoustics, Speech,
  and Signal Processing (ICASSP)}, vol.~6, May 2001, pp. 3997--4000.

\bibitem{Dahimene2008_declipping}
A.~Dahimene, M.~Noureddine, and A.~Azrar, ``A simple algorithm for the
  restoration of clipped speech signal,'' \emph{Informatica}, vol.~32, pp.
  183--188, 2008.

\bibitem{BilenOzerovPerez2015:declipping.via.NMF}
{\c{C}}.~Bilen, A.~Ozerov, and P.~P{\'e}rez, ``Audio declipping via nonnegative
  matrix factorization,'' in \emph{2015 IEEE Workshop on Applications of Signal
  Processing to Audio and Acoustics (WASPAA)}, Oct 2015, pp. 1--5.

\bibitem{Takahashi2015:Block.Adaptive.Decipping.Null.Space}
T.~Takahashi, K.~Uruma, K.~Konishi, and T.~Furukawa, ``Block adaptive algorithm
  for signal declipping based on null space alternating optimization,''
  \emph{IEICE Transactions on Information and Systems}, vol. E98.D, no.~1, pp.
  206--209, 2015.

\bibitem{Kitic2015:Sparsity.cosparsity.declipping}
S.~Kiti{\'c}, N.~Bertin, and R.~Gribonval, ``Sparsity and cosparsity for audio
  declipping: a flexible non-convex approach,'' in \emph{{LVA/ICA 2015} -- The
  12th International Conference on Latent Variable Analysis and Signal
  Separation}, Aug 2015, pp. 243--250.

\bibitem{ZaviskaRajmicSchimmel2019:Psychoacoustics.l1.declipping}
P.~{Z\'{a}vi\v{s}ka}, P.~{Rajmic}, and J.~{Schimmel}, ``Psychoacoustically
  motivated audio declipping based on weighted l1 minimization,'' in \emph{2019
  42nd International Conference on Telecommunications and Signal Processing
  (TSP)}, July 2019, pp. 338--342.

\bibitem{ZaviskaRajmicOzerovRencker2021:Declipping.Survey}
P.~Z{\'a}vi{\v s}ka, P.~Rajmic, A.~Ozerov, and L.~Rencker, ``A survey and an
  extensive evaluation of popular audio declipping methods,'' \emph{IEEE
  Journal of Selected Topics in Signal Processing}, vol.~15, no.~1, pp. 5--24,
  2021.

\bibitem{GaultierKiticGribonvalBertin:Declipping2021}
C.~{Gaultier}, S.~{Kiti\'c}, R.~{Gribonval}, and N.~{Bertin}, ``Sparsity-based
  audio declipping methods: selected overview, new algorithms, and large-scale
  evaluation,'' \emph{IEEE/ACM Transactions on Audio, Speech, and Language
  Processing}, vol.~29, pp. 1174--1187, 2021.

\bibitem{Adler2011:Declipping}
A.~Adler, V.~Emiya, M.~Jafari, M.~Elad, R.~Gribonval, and M.~Plumbley, ``A
  constrained matching pursuit approach to audio declipping,'' in
  \emph{2011 IEEE International Conference on Acoustics, Speech and Signal Processing (ICASSP)}, 
	May 2011, pp. 329--332.

\bibitem{Defraene2013:Declipping.perceptual.compressed.sensing}
B.~Defraene, N.~Mansour, S.~D. Hertogh, T.~van Waterschoot, M.~Diehl, and
  M.~Moonen, ``Declipping of audio signals using perceptual compressed
  sensing,'' \emph{IEEE Transactions on Audio, Speech, and Language
  Processing}, vol.~21, no.~12, pp. 2627--2637, 2013.

\bibitem{SiedenburgKowalskiDoerfler2014:Audio.declip.social.sparsity}
K.~Siedenburg, M.~Kowalski, and M.~Dorfler, ``Audio declipping with social
  sparsity,'' in \emph{2014 IEEE International Conference on Acoustics, Speech and Signal Processing (ICASSP)},
  May 2014, pp. 1577--1581.

\bibitem{RenckerBachWangPlumbley2018:Consistent.dictionary.learning.LVA}
L.~Rencker, F.~Bach, W.~Wang, and M.~D. Plumbley, ``Consistent dictionary
  learning for signal declipping,'' in \emph{{LVA/ICA 2018} -- The
  12th International Conference on Latent Variable Analysis and Signal
  Separation},
  July 2018, pp. 446--455.

\bibitem{Kabal2002:PEAQ}
P.~Kabal, ``An examination and interpretation of {ITU-R BS.1387}: Perceptual
  evaluation of audio quality,'' MMSP Lab Technical Report, Dept. Electrical \&
  Computer Engineering, McGill University, Tech. Rep., May 2002.

\bibitem{Huber:2006a}
R.~Huber and B.~Kollmeier, ``{PEMO-Q---A} new method for objective audio
  quality assessment using a model of auditory perception,'' \emph{IEEE Trans.
  Audio Speech Language Proc.}, vol.~14, no.~6, pp. 1902--1911, 2006.

\bibitem{siedenburg2012:wiener}
K.~Siedenburg, ``Persistent Empirical Wiener Estimation With Adaptive
  Threshold Selection for Audio Denoising,'' in \emph{Proceedings of
  the 9th Sound and Music Computing Conference}, July 2012, pp. 426--433.

\bibitem{DonoghueCandes2013:Adaptive.restart}
B.~O’Donoghue and E.~Candes, ``Adaptive restart for accelerated gradient schemes,''
  \emph{Foundations of Computational Mathematics}, vol.~15, no.~3, pp. 715--732, 2013.

\end{thebibliography}

\end{document}